\documentstyle [12pt]{article}
\newcommand{\doublespace}{
    \renewcommand{\baselinestretch}{1.6}\large\normalsize}

\newcommand{\beq}{\begin{equation}}
\newcommand{\eeq}{\end{equation}}
\newcommand{\ba}{\begin{eqnarray}}
\newcommand{\ea}{\end{eqnarray}}

\newcommand{\footar}{\footnote[2]{\small  Notice a misprint in the
formula for $\Pi^{\Delta h}$ in the reference \cite{Aou93}.
The results were produced with the correct ansatz.\\}}
\newcommand{\footju}{\footnote[1]{\small  also at: IKP (Theorie),
Forschungszentrum J\"ulich D-52425 J\"ulich, Fed. Rep. Germany\\}}

\begin{document}
\begin{titlepage}
\pagestyle{empty}
\begin{flushright}
{\small June, 1994}
%DRAFT}
\end{flushright}
\vspace{0.5cm}
\begin{center}
\doublespace
\begin{Large}
{\bf{S-Wave $\pi\pi$ Correlations in Dense Nuclear Matter}}
\end{Large}
\vskip 0.2in
Z. AOUISSAT$^{(1),}${\footju}, R. RAPP$^{(2),}$\footnotemark[1], \\
G. CHANFRAY$^{(3)}$, P. SCHUCK$^{(1)}$
and J. WAMBACH$^{(2),}$\footnotemark[1],\\
\vspace{0.2 in}
$^{(1)}$ {\small{\it  ISN, IN2P3-CNRS/Universit\'e Joseph Fourier\\
53 avenue des Martyrs, \\  F-38026 Grenoble C\'edex, France.}}\\
\vspace{0.2in}
$^{(2)}$ {\small{\it Department of Physics\\
University of Illinois at Urbana-Champaign\\
Urbana, IL 61801, U.S.A.}}\\
\vspace{0.2in}

$^{(3)}$ {\small{\it IPN-Lyon, IN2P3-CNRS/ Universit\'{e}
Claude Bernard LyonI,\\  43 Bd. du 11 Novembre 1918,\\
F-6922 Villeurbanne C\'{e}dex, France.}}\\

\end{center}
\vspace{1.5cm}

\begin{abstract}
The s-wave correlated $\pi$-$\pi$ strength distribution in cold nuclear
matter is investigated. The in-medium single pions are renormalized
by means of the usual p-wave couplings to nucleons and $\Delta$'s.
Two groups of models for the $\pi$-$\pi$ interaction are studied.
The first one uses phenomenological models
which reproduce the phase shifts but do not respect the soft-pion
theorems. With increasing density a strong accumulation of strength
in the subhreshold region is observed. Slightly above saturation
density there even occurs an instability with respect to s-wave
$\pi$-$\pi$ pair condensation. These effects are strongly
moderated or disappear altogether if constraints from chiral
symmetry are built into the $\pi$-$\pi$ interaction.

\end{abstract}
\end{titlepage}

\newpage
\pagestyle{plain}
\baselineskip 16pt
\vskip 48pt

\newpage
\doublespace
%%%%%%%%%%%%%%%%%%%%%%%%%%%%%%%%%%%%%%%%%%%%%%%%%%%%%%%%%%%%%%%%%%
%%%%%%%%%%%%%%%%%%%%%%%%%%%%%%%%%%%%%%%%%%%%%%%%%%%%%%%%%%%%%%%%%%
\section{Introduction}
Correlated two-pion exchange is a crucial piece of nucleon-nucleon
attraction at intermediate range \cite{Mach87}. On the other hand
pion-nucleus phenomenology implies that single-pion propagation is
strongly modified in a nuclear medium through couplings to
nucleons and $\Delta$'s \cite{ErWe}. It is therefore interesting
to ask to what extent pion-pion correlations are changed in a nuclear
medium. In a first investigation of this kind, Schuck et al.
\cite{SCN88} have shown that, indeed,
the influence of the medium on the $\pi$-$\pi$
correlations may be extremely strong in the scalar-isoscalar
($\sigma$-meson) channel. Subsequently also the vector-isovector
($\rho$-meson) channel was investigated \cite{Cha91} showing more
moderate effects. The strong influence
of the medium in the I=J=0 channel arises
from the fact that there is rather strong attraction between the pions
at low energies close to the $2m_{\pi}$ threshold. The in-medium
suppression of the single-pion kinetic energy through coupling to
$\Delta$'s and nucleons then leads to a threshold accumulation
of two-pion strength \cite{Cha91} or even the occurrence of
quasi-bound states. For moderate compression it was found in
\cite{Aou93}
that the system may become unstable with respect to $\pi$-$\pi$ pair
condensation. This instability is much softer than that for single-pion
condensation discussed long ago \cite{Mig50}. Such spectacular effects
call for more exhaustive investigations which are being carried out in
the present paper. We first repeat the calculation of \cite{Aou93}
using
a quite different, but very well tested, phenomenological $\pi$-$\pi$
interaction (the J\"{u}lich model \cite{Lohs}). This model also predicts
the instability (with slightly lower) critical density and thus
confirms the findings of \cite{Aou93}.

A dramatic reshaping of the $\pi$-$\pi$ strength distribution
inevitably would have strong and potentially disturbing effects on
the nuclear equation of state since it implies a very
long-range scalar interaction near the
critical density. One has therefore to look for physical effects
which possibly could oppose the softening of the $\pi$-$\pi$ mode. One
serious defect of phenomenological $\pi$-$\pi$ interactions is that
they often do not respect constraints from chiral symmetry
which are known to govern the near threshold behavior \cite{Wein67}.
 This may not be crucial
for energies well above the 2$m_{\pi}$ threshold but for subthreshold
effects it certainly is. We therefore investigate whether the
implementation of the soft-pion theorems can hinder an extreme
softening of the $\pi$-$\pi$ distribution. In a first attempt
we use the linear sigma model adjusting
the parameters such that the $I=0$ and $I=2$ scattering lengths as
well as the low-energy phase shifts are reasonably well reproduced.
Indeed it is found that this model yields a $\pi$-$\pi$ interaction
which, when continued
off-shell below the $2m_{\pi}$ threshold, builds up a strong repulsion.
The in-medium calculation of the $\pi$-$\pi$ distribution then shows
very little subthreshold strength and no pair condensation. The linear
sigma model does not include the $\rho$-meson which is a crucial
ingredient for the s-wave attraction of the J\"ulich model. To study
the influence of
the $\rho$-meson (under the constraints of chiral symmetry) we examine,
in a second step, the Lagrangian \cite{Wein67} of Weinberg, which
incorporates the $\rho$-meson as a gauge boson in the non-linear
$\sigma$ model. Also this model gives no significant
subthreshold strength, emphasizing the generic features of
chiral symmetry. The most crucial point
is the correct behavior of the scattering lengths in the chiral limit
$m_\pi \to 0$. We have therefore modified the J\"ulich model to impose
the correct chiral limits. This leads to the same conclusions suggesting
strongly that a $\pi$-$\pi$ s-wave pair instablity in dense
nuclear matter does not exist.

The paper is organized as follows: To set the stage in section 2 our
model for the renormalization of the single-pion propagator is outlined.
It includes p-wave pion-nucleon and pion-$\Delta$ interactions
with explicit treatment of the particle-hole continuum and the in-medium
$\Delta$ width. Section 3 is devoted to the calculation of s-wave
correlated
pions in nuclear matter by using the two phenomenological models of
Johnstone-Lee \cite{Lee} and the J\"ulich group \cite{Lohs}. Both
predict a pairing instability. In section 4 the constraints from chiral
symmetry are then included in the framework of
the linear sigma model, the gauged non-linear sigma model and an
extended version of the J\"ulich model, as discussed above.
Section 5 gives an overall discussion of our findings and concludes with
an outlook.

%%%%%%%%%%%%%%%%%%%%%%%%%%%%%%%%%%%%%%%%%%%%%%%%%%%%%%%%%%%%%%%%%%%
\section{Single-Pion Propagator}
%%%%%%%%%%%%%%%%%%%%%%%%%%%%%%%%%%%%%%%%%%%%%%%%%%%%%%%%%%%%%%%%%%%
Let us begin with a brief description of the model for renormalizing
the single-pion propagator in nuclear matter. Including nucleons and
$\Delta$-isobars as the most important degrees of freedom and using
p-wave dominance in $\pi N$ scattering the system is
described by the Hamiltonian
\beq
H=T_\pi+T_N+T_\Delta+V^N_{\pi N}+V^N_{\pi\Delta}+V^{NN}_{NN}+
V^{NN}_{N\Delta}+V^{NN}_{\Delta\Delta}  \ ,
\eeq
where $T_\pi,T_N$ and $T_\Delta$ denote the kinetic energy operators and
$V^N_{\pi N},V^N_{\pi\Delta}$ are the $\pi NN$ and $\pi N\Delta$
interactions, respectively. Short-range
interactions among the nucleons and isobars are described by
the potentials $V^{NN}_{NN}, V^{NN}_{N\Delta}$ and
$V^{NN}_{\Delta\Delta}$.
The in-medium pion propagator D$_\pi$ can then be
extracted from a set of coupled-channel integral equations (Fig.~1)\\
\ba
D_{\pi} & = & D_{\pi}^{0}+D_{\pi}^{0}V_{\pi N}^{N}\tau_{N\pi}^{N}+
D_{\pi}^{0}V_{\pi\Delta}^{N}\tau_{\Delta\pi}^{N} \ ,
\nonumber\\
\tau_{N\pi}^{N} & = & \chi_{N}^{N}V_{N\pi}^{N}D_{\pi}+
\chi_{N}^{N}V_{NN}^
{NN}\tau_{N\pi}^{N}+\chi_{N}^{N}V_{N\Delta}^{NN}\tau_{\Delta\pi}^{N} \ ,
\nonumber\\
\tau_{\Delta\pi}^{N} & = & \chi_{\Delta}^{N}V_{\Delta\pi}^{N}D_{\pi}+
\chi_
{\Delta}^{N}V_{\Delta\Delta}^{NN}\tau_{\Delta\pi}^{N}+\chi_{\Delta}^{N}
V_{\Delta N}^{NN}\tau_{N\pi}^{N} \ ,
\ea
where $D_{\pi}^{0}$ denotes the bare pion propagator
\beq
D_{\pi}^{0}(\omega, q)=\frac{1}{\omega^{2}-\omega^{2}_q+i\eta}
\eeq
with $\omega_q^2=q^2+m_\pi^2$,
while $\chi_{N}^{N}$ and $\chi_{\Delta}^{N}$ are the particle-hole and
$\Delta$-hole Lindhard functions, respectively.
Parametrizing the fermionic interaction potentials by Migdal parameters
it is then straightforward to cast the pion propagator in the following
form:
\beq
D_{\pi}={1\over\omega^2-\omega_q^2- \Sigma_\pi(\omega,q)+i\eta}
\eeq
with the self energy given by
\beq
\Sigma_\pi(\omega,{\bf q})=  -q^2{\Pi^{ph}+\Pi^{\Delta h}+
\Pi^{ph}\Pi^{\Delta h}(2g'_{N\Delta}-g'_{NN}-g'_{\Delta\Delta})
\over(1+g'_{NN}\Pi^{ph})(1+g'_{\Delta\Delta}\Pi^{\Delta h})-
g'^2_{N\Delta}\Pi^{ph}\Pi^{\Delta h}} \ ,
\eeq
where
\beq
\Pi^{ph}(\omega,{\bf q})=-4\frac{f^{2}_{\pi}\Gamma^2_\pi(q)}
{m_{\pi}^{2}}
i\int\frac
{d^{4}p}{(2\pi)^4}G^{0}_{N}(p)G^{0}_{N}(p+q)
\eeq
and
\beq
\Pi^{\Delta h}(\omega,{\bf q})=-\frac{16}{9}\frac{f^{\ast^{2}}_{\pi}
\Gamma^2_\pi(q)}
{m_{\pi}^{2}}i\int\frac{d^{4}p}{(2\pi)^4}G^{0}_{N}(p)
 G^{0}_{\Delta}(p+q)
\eeq
with $\Gamma_\pi(q)$ being the usual monopole-type form factor at the
$\pi$NN and $\pi$N$\Delta$ vertices with a cutoff of 1.2GeV/c.
Using the nucleon and $\Delta$ Green's functions, $G_{N}^{0}$ and
$G_{\Delta}^{0}$, the integrals in eqs.~(2.6) and (2.7)  can  be easily
expressed in terms of the $ph$ and
$\Delta h$ Lindhard functions \cite{ErWe,Fet71}.
Due to the high-energy cut in the $\Delta$-h bubbles, Fermi motion
effects in eq.~(5) can be ignored (which we have verified numerically),
and we can put $\Pi^{\Delta h}$ in the simple form{\footar}
\beq
\Pi^{\Delta h}(\omega,q)=-\frac{4}{9}\frac{f^{\ast^{2}}_{\pi}
\Gamma^2_\pi(q)}
{m_{\pi}^{2}}\rho\left\{\frac{1}{\omega-\epsilon_{\Delta h}({\bf q})
+i\frac{1}{2}\Gamma_{\Delta}}-\frac{1}{\omega+\epsilon_{\Delta h}
({\bf q})} \right\}
\eeq
with
\beq
\epsilon_{\Delta h}(q)=M_{\Delta}-M_{N}+\frac{q^2}{2M_{\Delta}} \ .
\eeq
To account for the finite life time of the $\Delta$-isobar, one has to
consider a complex $\Delta$-energy with an in-medium $\Delta$-decay
width, $\Gamma_\Delta$.
In this study, we shall include both Pauli blocking
and relativistic corrections as well as a term which accounts for
higher-order in-medium corrections such that
\beq
\Gamma_{\Delta}=\Gamma_{\Delta}^{(1)}+\Gamma_{\Delta}^{(2)}.
\eeq
$\Gamma_{\Delta}^{(1)}$ is taken from reference \cite{Oset87}. For
$\Gamma_{\Delta}^{(2)}$ we adopt the following phenomenological ansatz:
\beq
\Gamma_{\Delta}^{(2)}=48\frac{\rho}{\rho_0}\left(1+\frac{\omega-m_\pi}
{380-m_\pi}\right)\Lambda(\omega),
\eeq
where $\Lambda(\omega$) is a cutoff function, which suppresses
$\Gamma_{\Delta}^{(2)}$ smoothly below $m_\pi$ and above some 400 MeV.
This choice is motivated by the observation that the inelastic part
of the $\Delta$-self energy in the two processes: $\Delta\rightarrow
\pi N$ and $\Delta\rightarrow\gamma N$ turns out to be very similar.
This seems to indicate that $\Gamma_{\Delta}^{(2)}$
is insensitive to the nature of emitted particles other than the nucleon
and one is encouraged to use it for off-shell pions as well. Although
somewhat rough, the ansatz (11) aims to account for the
well-known $\rho^{2}$ behavior in the inelastic part of the
$\pi$-nucleus optical potential \`{a} la Kisslinger \cite{Kis55}.
It is also important to keep in mind, however, that this contribution
is not sufficient to reproduce quantitatively the absorptive
part in the optical
potential. One therefore has to consider also contributions from the
2p-2h excitations induced by the pion.
Following the exhaustive analysis of the
Lyon-group \cite{Alb84,Cha84}, we adopt the following phenomenological
expression for the imaginary part of the 2p-2h pion self energy:
\beq
Im\Pi^{2p-2h}(\omega,q)=-0.3\Gamma_\pi^{2}({\bf q})
(\frac{\rho}{\rho_0})^2\Lambda(\omega) \ .
\eeq
The coefficient 0.3 is fitted such that the combined effects of
$\Gamma_{\Delta}^{(2)}$ in the RPA approximation and the 2p-2h
contribution properly accounts for the empirical absorptive part of the
$\pi$-nucleus optical potential at threshold.
The real part of the 2p-2h pion self energy, although straightforward to
evaluate via dispersion relations, will not be considered since its
influence is minor in the present study. For practical purposes we
absorb $\Pi^{2p-2h}$ into the $\Delta$-hole polarization propagator as
\beq
\Pi^{\Delta h}\to \Pi^{\Delta h} + \Pi^{2p-2h} .
\eeq
This concludes the outline of our in-medium renormalization of the pion
propagator.
%%%%%%%%%%%%%%%%%%%%%%%%%%%%%%%%%%%%%%%%%%%%%%%%%%%%%%%%%%%%%%%%%%%%%%
%%%%%%%%%%%%%%%%%%%%%%%%%%%%%%%%%%%%%%%%%%%%%%%%%%%%%%%%%%%%%%%%%%%%%%
\section {$\pi$-$\pi$ Correlations in Nuclear Matter:
Phenomenological Models}
%%%%%%%%%%%%%%%%%%%%%%%%%%%%%%%%%%%%%%%%%%%%%%%%%%%%%%%%%%%%%%%%%%%%%%
\subsection{The Johnstone-Lee Model}
To study the $\pi\pi$ correlations in dense nuclear matter, we
first introduce a simple phenomenological model for the scattering of
two pions in vacuum. This model is based on the separable potentials of
Johnstone and Lee \cite{Lee}. For two-pion CMS energies below $\sim 1$
GeV these give a realistic description of the $\pi\pi$ phase-shifts
in low partial waves. Aside from numerical advantages, the separability
renders simple and transparent expressions for the $\sigma$ propagator.
In the next section we shall use the more sophisticated
meson-exchange model by the J\"ulich group, which is able to extend the
description above the $K\bar K$ threshold.

In the model of Johnstone and Lee, the interaction
kernel for $\pi\pi$ scattering in the $J^P=0^+,I=0$ channel is of the
form
\beq
V_{\pi\pi}(q,q',E)=2\pi^{2}g^{2}_{\sigma}2\omega_{q}2\omega_{q'}
v_{\sigma}(q)D_{\sigma}^{0}(E)v_{\sigma}(q'),
\eeq
where
\beq
v_{\sigma}(q)=\sqrt{2M_{\sigma}}/(1+(\frac{q}{q_\sigma})^2)
\eeq
denotes the finite-size $\sigma\pi\pi$ vertex and
\beq
D_{\sigma}^{0}(E)=\frac{1}{E^{2}-M_{\sigma}^{2}+i\eta}
\eeq
the bare $\sigma$-propagator. The coupling constant $g_\sigma$, the
cutoff $q_\sigma$ and
the bare $\sigma$-mass $M_\sigma$ are free parameters, adjusted to the
experimental $I$=0-, s-wave phase shift and the scattering length.
The T-matrix in "back to back" kinematics then takes
a very simple form
\beq
T^{00}_{\pi\pi}(q,q',E)=2\pi^{2}g^{2}_{\sigma}2\omega_{q}2\omega_{q'}
 v_{\sigma}(q)D_{\sigma}(E)v_{\sigma}(q'),
\eeq
which is the same as (14) except that $D_\sigma^0$ is replaced
by the dressed propagator
\beq
D_{\sigma}(E)=\frac{1}{E^{2}-M_{\sigma}^{2}-\Sigma_{\sigma}(E)+i\eta}
 \ .
\eeq
Here $\Sigma_\sigma$ is the $\sigma$-meson self energy due to $\pi\pi$
rescattering processes involving s-channel pole graphs. Explicitly
\beq
-i\Sigma_{\sigma}(E)=2\pi^{2}g_{\sigma}^{2}\int{d\omega d^{3}q\over
(2\pi)^4}4\omega_{q}^{2}v_{\sigma}^{2}(q)D^0_{\pi}(\omega,q)
D^0_{\pi}(E-\omega,q) \ .
\eeq
In nuclear matter, modifications of the T-matrix enter
through the dressing of the pion with $ph$ and $\Delta h$
excitations (Fig.~2). This amounts to replacing
$D^0_\pi$ with the renormalized propagator $D_\pi$ given in sect.~2.
In previous work \cite{Cha91}, we have used a simple two-level model
by coupling  the pion only to the $\Delta$-hole branch. As a result,
a strong accumulation of strength near the 2$m_\pi$ threshold was found
in the  sigma-mass distribution. For densities somewhat above saturation
density even a two-pion bound state emerges. The shift of the $\sigma$
strength to low energy can be easily  understood as being due to the
softening of the pionic branch induced by the well-known level-mixing
mechanism.
In the present paper we can go beyond this simplified treatment by
incorporating the particle-hole branch, including Fermi motion. It is
clear that any strength accumulation below the two-pion
threshold will depend sensitively on the energy
distribution of the $ph$ excitations, thus Fermi
motion and Pauli blocking effects are crucial. Secondly we include
a realistic treatment of the $\Delta$-decay width as well as
absorptive  effects as described in sect.~2. With these extensions the
 energy integration in eq.(19) can no longer be done analytically
as it is the case in the pure two-level model of \cite{Cha91}. Obviously
a numerical integration over the entire range of energy is not a trivial
task. To overcome this difficulty we first compute
the imaginary part of $\Sigma_\sigma$ given by:
\beq
Im\Sigma_{\sigma}(E)=g_{\sigma}^{2}\int dq q^{2}4\omega^{2}_{q}
v_{\sigma}^{2}(q) ImG_{\pi\pi}(E,q)
\eeq
with $G_{\pi\pi}$ being the two-pion propagator
(the convolution of two single-pion propagators). Its imaginary part
can be computed by using the following expression \cite{Aou93,Cha93},
\beq
ImG_{\pi\pi}(E,q)=-\frac{1}{\pi}\int_{0}^{E} d\omega ImD_{\pi}
(\omega,q)ImD_{\pi}(E-\omega,q) \ ,
\eeq
which simplifies the numerical integration considerably since
off-shell integrations can be avoided. The real part of $\Sigma_\sigma$
is computed by means of the usual dispersion relation, that is
\beq
Re\Sigma_{\sigma}(E)=\frac{1}{\pi}{\cal P}\int_{0}^{+\infty}dE'^{2}
\frac{Im\Sigma_{\sigma}(E')}{E'^2-E^2} \ ,
\eeq
where ${\cal P}$ denotes the Cauchy principal value.
To obtain results we have to specify the Fermi liquid parameters,
which represent the short-range repulsion in the
nucleon and $\Delta$ interaction potentials as discussed in sect.~2.
We have chosen the following values:
\beq
 \begin{array}{l}
   g'_{NN}=0.8,\\
  g'_{N\Delta}=g'_{\Delta\Delta}=0.5
 \end{array}
\eeq
based on empirical knowledge from Gamow-Teller systematics \cite{Ost}
as well as information from pion double-charge exchange reactions
\cite{John}. In Fig.~3 we contrast results from the simple
two-level model with those including isobar-decay and absorption
effects. As expected, these modifications
lead to a broadening of the bound state but have a minor effect on the
overall distribution. The situation changes if the particle-hole branch
is included. In this case the two-pion state couples to the
particle-hole continuum, resulting in an enhancement of the decay
width below threshold as can be seen from the upper panel in Fig.~4,
where the imaginary part of $\Sigma_\sigma$ is displayed.
The width strongly increases  with density.In the real
part of $D_\sigma$ (lower panel in Fig.~4) one observes a rapid
decrease of the
2$\pi$ pole towards zero energy, signalling the pairing
instability. The critical density is slightly above 1.3 $\rho_0$.
This is significantly lower than the threshold for pion condensation,
which  in our
model occurs at $3.6\rho_0$. The reason is the
extra attraction provided by the s-wave $\pi\pi$ interaction.
The approach to pair condensation is also seen
clearly in the $\sigma$-strength function (Fig.~5)
where we notice that, with
increasing density, a resonance below threshold develops which becomes
narrower and absorbs practically the entire strength close to the
instability.

%%%%%%%%%%%%%%%%%%%%%%%%%%%%%%%%%%%%%%%%%%%%%%%%%%%%%%%%%%%%%%%%%%%%%%
\subsection{The J\"ulich Model}
%%%%%%%%%%%%%%%%%%%%%%%%%%%%%%%%%%%%%%%%%%%%%%%%%%%%%%%%%%%%%%%%%%%%%%
In order to check to what extent the results depend  on
the specific form of the $\pi\pi$ interaction we have repeated the
calculations for the model of Lohse et al. \cite{Lohs}.
Here $V_{\pi\pi}$ is constructed from meson-exchange including the most
important {\it s}- and {\it t}-channel contributions. Besides the
$\sigma$ pole term, the {\it t}-channel $\rho$ exchange is very
important for s-waves (Fig.~6). In addition,
the model couples  the $K\bar K$ channel,
which becomes crucial for energies $\geq 1$ GeV. To derive an equation
for the $T$-matrix, a Blankenbecler-Sugar (BbS)
reduction \cite{BbS}  of the 4-dimensional Bethe-Salpeter equation is
employed maintaining covariance \cite{PeHS}.
The $T$-matrix for given angular momentum $J$ and
isospin $I$ is then obtained as
\beq
T_{\mu\nu}^{JI}(E;q,q') = V_{\mu\nu}^{JI}(E;q,q')+
%    \nonumber\\
%& &
+\sum_{\lambda =1}^{2} \  \int_0^{\infty}dk \ k^2 \ 4{\omega}_k^2 \
V_{\mu\lambda}^{JI}(E;q,k)
 \   G_{\lambda}^0(E,k) \ T_{\lambda\nu}^{JI}(E;k,q')  \ ,
\eeq
where the indices $\mu,\nu,\lambda$=1,2 label the two
different channels $\pi\pi$ or $K\bar K$, respectively (also
$\omega_k^2=m_\lambda^2+k^2$ with $m_\lambda=m_\pi,m_K$).
In the BbS form the vacuum $\pi\pi$-/$K{\bar K}$-propagator is
given by
\beq
G_{\lambda}^0(E,k)=\frac{1}{\omega_k} \
\frac{1}{E^2-4\omega_k^2+i\eta} \ .
\eeq
This model gives a good description of the phase shifts and
inelasticities
up to $\sim 1.5$ GeV.

For the present problem it is most relevant that the J\"ulich model has
a different off-shell behavior than the Johnstone-Lee model which
could potentially alter the subthreshold behavior of $D_\sigma$ since
the pions are far off shell.

Because of the $t$-channel interaction the calculation of the $\sigma$
propagator is more involved than in the separable model.  The way we
proceed is as follows:
in a first step we calculate the full off-shell T-matrix with the
two pions embedded in nuclear matter.
In analogy to sect. 3, we make use of a
dispersion relation, i.e. we calculate $Im G_{\pi\pi}$ using
eq.~(21) and evaluate the real part as
\beq
Re G_{\pi\pi}(k,E)=\frac{1}{\pi}{\cal P} \int_{0}^{+\infty} dE'^2 \
\frac{ImG_{\pi\pi}(k,E')}{E'^2-E^2} \ .
\eeq
The T-matrix equation is then solved numerically with the replacement
$G^0_{\pi\pi}  \to G_{\pi\pi}$ (the $K\bar K$ propagator remains
unmodified)
employing the Haftel-Tabakin method~\cite{HaTa}. The in-medium T-matrix,
${\tilde T}_{\mu\nu}^{00}$, then serves to dress the bare
$\sigma$-propagator given by eq.~(16) with a bare mass
 of $M_\sigma=2.2$ GeV (in free space the renormalization of
$D_\sigma^0(E)$, generated by the iteration of $V$, leads
to the $f_0(1400)$-resonance). The medium-modified $\sigma$-propagator
then takes the following  form
\ba
D_\sigma(E) = D_\sigma^0(E) + & \!\!\!\!D_\sigma^0(E) & \!\!\!\lbrack
\sum_{\lambda =1}^{2}
\int_{0}^{\infty} dq q^2 4\omega_q^2  v^\lambda_\sigma(E,q) \
G_\lambda(E,q) \ v^\lambda_\sigma(E,q)
\nonumber\\
 & & \!\!\!\!\!\!\! + \sum_{\mu =1}^{2} \int_{0}^{\infty} dq q^2
 4\omega_q^2
 v^\mu_\sigma(E,q) \ G_\mu(E,q) \
\nonumber\\
 & & \!\!\!\!\!\!\! \times\sum_{\nu =1}^{2} \int_{0}^{\infty} dq' q'^2
 4\omega_{
q'}^2
{\tilde T}_{\mu\nu}^{00}(E;q,q') \ G_\nu(E,q') \ v^\nu_\sigma(E,q')
\rbrack
\ D_\sigma^0(E)  \  ,
\nonumber\\
\ea
where $v^\lambda_\sigma(E,q)$ denote the vertex functions for the
s-channel transitions $\pi\pi, \ K{\bar K} \rightarrow \sigma$. The
second term on the right hand side of eq.~(28) accounts for the
single $\pi\pi$-/$K{\bar K}$-bubble, whereas the third term sums all
possible $\pi\pi$- and $K{\bar K}$-rescattering processes. We should
emphasize that it is necessary to include the
contributions from the $K{\bar K}$ channel which, off-shell, enters
below the $K\bar K$ threshold at $2m_K$. Neglecting this channel results
in an unphysical pole at the bare mass $M_\sigma$.
For energies $E>2m_K$ we have introduced a small width for the kaons
to avoid numerical instabilities (moderate changes in the width
parameter affect our results for $D_\sigma(E>2m_K)$ only slightly;
this region is of minor interest in the present discussion anyway).
Fig.~7 displays the results for density values $\rho/\rho_0=0.5,1.0$ and
1.2. They are very similar to those obtained with the separable model
(Fig.~5). The critical density for pairing is somewhat lower at
$\rho/\rho_0\approx 1.3$, which can be traced back to the somewhat
larger
s-wave attraction in the J\"ulich model. This result seems to suggest
that any $\pi\pi$ model, which reproduces the phase shifts, leads
to an instability at densities slightly above saturation density.

%%%%%%%%%%%%%%%%%%%%%%%%%%%%%%%%%%%%%%%%%%%%%%%%%%%%%%%%%%%%%%%%%%%%%
%%%%%%%%%%%%%%%%%%%%%%%%%%%%%%%%%%%%%%%%%%%%%%%%%%%%%%%%%%%%%%%%%%%%%
\section{$\pi$-$\pi$ Correlations in Nuclear Matter: Chiral Models}
%%%%%%%%%%%%%%%%%%%%%%%%%%%%%%%%%%%%%%%%%%%%%%%%%%%%%%%%%%%%%%%%%%%%%
As mentioned in the introduction, the models used in the previous
section do not respect the constraints from chiral symmetry, i.e.
the soft pion theorems which imply a linear decrease of the s-wave
$\pi$-$\pi$ $I=0$ and $I=2$ scattering lengths as
$m_{\pi}\rightarrow 0$.
We will now elaborate in more detail on these properties and then
apply various chiral models below.

 It is possible to relate the I=J=0 $\pi$-$\pi$
scattering amplitude at the soft point, defined through the
Mandelstam variables $s=m_{\pi}^{2}, t=0, u=m_{\pi}^{2}$, to the
$\pi$-$\pi$ sigma commutator $\Sigma_{\pi\pi}$ as
\beq
M^{00}(s=m_{\pi}^{2}, t=0, u=m_{\pi}^{2}) = \frac{\Sigma_{\pi\pi}}
{f_{\pi}^2},
\eeq
with $\Sigma_{\pi\pi}$ given by
\beq
\Sigma_{\pi\pi}=<\pi^{i}({\bf k=0})|[Q_{5},[Q_{5},H]]
|\pi^{i}({\bf k=0})>_{connected},
\eeq
where $Q_{5}$ is the axial charge and H the full strong-interaction
Hamiltonian. The (discrete) normalization is such that
\beq
<\pi^{i}({\bf k})|\pi^{i}({\bf k'})>= 2\omega_{{\bf k}}
\delta_{{\bf kk'}}.
\eeq
The derivation of this result only requires the standard reduction
formula
and PCAC and follows exactly the same lines as the derivation of the
$\pi$-nucleon sigma term (see for instance standard text books such as
\cite{Cheng}). An explicit calculation shows that the double commutator
appearing in eq.~(29) is just the piece of the Hamiltonian which
explicitly
breaks $SU(2)\times SU(2)$ chiral symmetry. Hence, in an effective
theory with pions, this yields
\beq
\Sigma_{\pi\pi}= \int d^{3}{\bf x} \frac{1}{2} m_{\pi}^{2}
< \pi ({\bf k=0})|:\Phi_{i}^{2}({\bf x},t):|\pi ({\bf k=0})> =
m_{\pi}^{2}.
\eeq
where $\Phi_{i}$ is the canonical pion field with isospin i. We thus
obtain a low-energy theorem
\beq
M^{00}(s=m_{\pi}^{2}, t=0, u=m_{\pi}^{2}) = \frac{m_{\pi}^2}
{f_{\pi}^2}.
\eeq
The reader may notice that this result is valid up to $m_{\pi}^{4}$
corrections as it is implicit in PCAC. Eq.~(32) should be
contrasted with the Weinberg result \cite{Wein67} at the physical
threshold which gives
\beq
M^{00}(s=4m_{\pi}^{2}, t=0, u=0) =- 7\frac{m_{\pi}^2}{f_{\pi}^2}.
\eeq
and indicates a sign change when going off-shell. However, the
soft-point amplitude is not directly relevant to the $\pi$-$\pi$
scattering amplitude where the Mandelstam variables are to be chosen as
\ba
s & = & E^2  ,\nonumber\\
t & = & 2m_{\pi}^{2}-2\omega_{q}\omega_{q'}+2{\bf qq'},\nonumber\\
u & = & 2m_{\pi}^{2}-2\omega_{q}\omega_{q'}-2{\bf qq'}\,
\ea
with {\bf q} and {\bf q}' being the relative momenta of the in- and
outgoing pion in the CM frame. In the medium it will be possible to
reach kinematical conditions which are relatively close to the point
where the soft pion result applies. For instance at
${\bf q=q'=}0$ and $E=m_{\pi}$ one has $s=m_{\pi}^{2}, t=u=0$.
Hence, one may suspect that the soft-pion
constraint will generate repulsion below the threshold which is
completely absent in the phenomenological models of Johnstone-Lee
and the J\"ulich group.

%%%%%%%%%%%%%%%%%%%%%%%%%%%%%%%%%%%%%%%%%%%%%%%%%%%%%%%%%%%%%%%%%%%%%%%
\subsection{ The Linear Sigma Model}
%%%%%%%%%%%%%%%%%%%%%%%%%%%%%%%%%%%%%%%%%%%%%%%%%%%%%%%%%%%%%%%%%%%%%%%
The linear $\sigma$ model does obey the soft-pion theorem
eq.~(32). At the tree level, the I=0 amplitude is given
by the Born amplitude ($m_\sigma$ is the mass of the $\sigma$ meson)
\beq
M^{I=0}_{B}(s, t, u) = \frac{m_{\sigma}^{2}-m_{\pi}^{2}}{f_{\pi}^2}
(3 \frac{s-m_{\pi}^{2}}{s-m_{\sigma}^2}
+ \frac{t-m_{\pi}^{2}}{t-m_{\sigma}^2}
+ \frac{u-m_{\pi}^{2}}{u-m_{\sigma}^2})
\eeq
which satisfies
\beq
M^{00}(s=m_{\pi}^{2}, t=0, u=m_{\pi}^{2}) = \frac{m_{\pi}^{2}}
{f_{\pi}^2}(1- \frac{m_{\pi}^{2}} {m_{\sigma}^{2}}).
\eeq
We also see that, for $s=m_{\pi}^{2}, u=t=0$, which can be
reached in subthreshold $\pi$-$\pi$ scattering, the amplitude is
even more repulsive, namely
\beq
M^{00}(s=m_{\pi}^{2}, t=0, u=0) = 2 \frac{m_{\pi}^{2}}
{f_{\pi}^2}(1- \frac{m_{\pi}^{2}} {m_{\sigma}^{2}}).
\eeq
We will now use the Born amplitude (35) of the linear $\sigma$ model
to construct a seperable $\pi$-$\pi$ potential. With the kinematical
choice (34) for the Mandelstam variables,
neglecting the t- and u-dependence in the denominators
(t,u $\ll m_\sigma$) and adding phenomenological form factors
to regularize the momentum integrals, we have
\beq
V^{00}(E,q,q') = \frac{m_{\sigma}^{2}-m_{\pi}^{2}}{f_{\pi}^2}
(3 \frac{E^{2}-m_{\pi}^{2}}{E^{2}-m_{\sigma}^2}
+ \frac{4\omega_{q}\omega_{q'}-2m_{\pi}^{2}}{m_{\sigma}^2})v(q)v(q') \ .
\eeq
The vacuum s-wave $\pi \pi$ amplitude is obtained by solving the
Lippmann-Schwinger equation
\beq
M^{00}(E,q,q') = V^{00}(E,q,q) + \int \frac{d^{3}{\bf k}}
{(2\pi)^3} \frac{ V^{00}(E,q,k) T^{00}(E,k,q')}
{2\omega_{k}(E^{2}-4\omega^2_{k}+i\eta)}
\eeq
With
\beq
m_{\sigma}=1.0  {\rm GeV},\qquad
v(q)=(1+ \frac{q^2}{64 m_{\pi}^2})^{-1}
\eeq
the scattering length is $a^{00} = 0.23 m_{\pi}^{-1}$ and we obtain
a good fit to the phase shift up to $E\approx 900$ MeV (Fig.~8).

\noindent
As before, the in-medium scattering matrix is evaluated by
modifying the 2-$\pi$ propagator according to eqs.~(21) and (26).
Rather than displaying the $\sigma$ propagator, as in sect.~3
we here show the invavariant amplitude $M^{00}$ as in (39) because
for some of the chiral models a $\sigma$ propagator cannot be
defined. However, for the non-chirally symmetric models of chapter 3
the imaginary part of the $M^{00}$ amplitude has the same shape as
$Im D_\sigma$  as displayed in Figs.~5 and 7.
As can be seen from Fig.~9, the density-dependence of
the strength distribution (upper panel) is rather weak. On the other
hand, the real part (lower panel) changes sign at low energy
reflecting the repulsive character of the $\pi\pi$ potential (38) close
to zero energy. This behavior prevents the pairing instability.

%%%%%%%%%%%%%%%%%%%%%%%%%%%%%%%%%%%%%%%%%%%%%%%%%%%%%%%%%%%%%%%%%%%%%%%%%
\subsection{The Weinberg Lagrangian}
%%%%%%%%%%%%%%%%%%%%%%%%%%%%%%%%%%%%%%%%%%%%%%%%%%%%%%%%%%%%%%%%%%%%%%%%%
To study the influence of the $\rho$-meson we now proceed to the
non-linear realization of chiral symmetry as proposed by Weinberg
\cite{Wein67}. Before going to nuclear matter, let us first
briefly recall the model. In the pure mesonic sector, the Lagrangian
can be cast in the following form
\beq
{\cal L}_{W}=-\frac{1}{4}F_{\mu \nu}F^{\mu \nu} - \frac{1}{2}
\frac{\partial_{\mu}{\bf \pi}\cdot \partial^{\mu}{\bf \pi}}
{(1+\frac{{\bf \pi}^2}
{4f_{\pi}^{2}})^2}- \frac{1}{2}m_{\pi}^{2}
\frac{{\bf \pi}^2}{1+\frac{{\bf \pi}^2}{4f_{\pi}^{2}}}
- \frac{1}{2}m_{\rho}^{2}[\rho_{\mu} + \frac{g_{\rho\pi\pi}}
{m_{\rho}^{2}} \frac{{\bf \pi}\times \partial_{\mu}
{\bf \pi}}{1+\frac{{\bf \pi}^2}{4f_{\pi}^{2}}}]^{2}.
\eeq
Along with the pions, the $\rho$-meson is introduced via gauging,
maintaining chiral symmetry. Here,
${\bf \rho}_{\mu}, F_{\mu \nu}$,and $m_{\rho}$ denote
the canonical field, the field strength tensor and the $\rho$-mass,
respectively. The coupling constant $g_{\rho \pi\pi}$
is fixed by the KSFR relation\cite{KSFR}
\beq
2  g_{\rho \pi\pi}^{2}f_{\pi}^{2}=m_{\rho}^{2} \ .
\eeq
To lowest order in the expansion of the pion field
one can derive the following Born amplitude from
eq.~(41)
\beq
M_B^{I=0}=M^{I=0}_{1} + M^{I=0}_{2} + M^{I=0}_{3},
\eeq
with
\ba
M^{I=0}_{1}(s, t, u) & = & - \frac{1}{f_{\pi}^2}
[3 (s-m_{\pi}^{2})+(t-m_{\pi}^{2})+ (u-m_{\pi}^{2})] \\
M^{I=0}_{2}(s, t, u) & = & \frac{2}{f_{\pi}^2}[(s-u) + (s-t)] \\
M^{I=0}_{3}(s, t, u) & = & \frac{2m_{\rho}^{2}} {f_{\pi}^2}
[ \frac{s-u}{t-m_{\rho}^2}+ \frac{s-t}{u-m_{\rho}^2}]
\ea
In the above expressions we have used the usual metric in which
$g^{00}=1, g^{ij}=-1$. The amplitude $M^{I=0}_{1}$ arises from
the contact terms induced by the covariant derivative and the symmetry
breaking term of the Lagrangian (2nd and 3rd term on the r.h.s.
of eq.~(41)). The other two amplitudes follow  from
the $\rho$-meson exchange and the contact term arising naturally
from the gauging procedure (last term on the r.h.s. of eq.~(41)).
As pointed out by Weinberg, $M^{I=0}_{2}$ and $M^{I=0}_{3}$ cancel
at the physical threshold such that only the
amplitude $M^{I=0}_{1}$ contributes to the scattering length.
One obtains $a^{00}=0.17 m_\pi^{-1}$ as predicted by Current-Algebra
\cite{Wein66}.

To go beyond the tree level we supplement, as for the linear sigma
model, the Born amplitude (43) with phenomenological form factors
of dipole type,
\beq
F(q)=\lbrack\frac{\Lambda^2+4m_\pi^2}{\Lambda^2+4\omega_q^2}\rbrack^2 .
\eeq
In contrast to the previous section,
where the on-mass-shell identification of the kinematic
variables has been employed, we choose the Blankenbecler-Sugar
prescription \cite{BbS} (also used in J\"ulich model) to reduce the
4-dimensional Bethe-Salpeter equation. In this case the two pions in
the intermediate state are put on-energy shell. It
implies the following identification of the Mandelstam variables
\ba
s & = & (q_1+q_2)^2=E^2 \ ,                     \nonumber \\
t & = & (q_1-q_3)^2=-(\vec {q} -\vec {q'})^2 \ ,  \nonumber \\
u & = & (q_1-q_4)^2=-(\vec {q} +\vec {q'})^2 \ .
\ea
which then determine the energy-momentum dependence
of the pseudopotentials $V(E,q,q')$ suitable for iteration.

With form factor cutoff values $\Lambda_1\approx$ 800 MeV,
 $\Lambda_2\approx$ 1300 MeV and $\Lambda_3\approx$ 2000 MeV
we can approximately reproduce the experimental s-wave scattering
phase shifts up to $E\approx 900MeV$.

As can be easily verified, the pseudopotential (38) of the linear sigma
model preserves the correct $m_\pi$ dependence of the scattering length.
With the kinematical choice (48) this is no longer the case. This
is partly due to the fact that the pseudopotential is no longer
separable. The scattering lengths $a^{00}$ and $a^{02}$ actually
diverge in the chiral limit. Furthermore, the repulsion in the
subthreshold region, which is present at the tree level,
does not survive the iteration. In medium this implies
again an s-wave instability for slightly compressed nuclear matter.

Thus, inspite of the special care given to chiral constraints at the
tree level, one must be cautious when iterating. There is, however,
a way of preserving the chiral limit to all orders. One employs
a subtracted dispersion relation to extract the real part of
$G_{\pi \pi}$  while preserving unitarity.
We start from the spectral representation of the 2-pion
propagator
\begin{equation}
G_{\pi\pi}(s,k)=-\frac{1}{\pi} \int ds' \ \frac{Im G{\pi\pi}(s',k)}
{s-s'+i\eta} \ .
\end{equation}
In vacuum,  the imaginary part reads
\begin{equation}
Im G_{\pi\pi}(s,k)=-\frac{\pi}{\omega_k} \ \delta (s-4\omega_k^2) \ .
\end{equation}
In principle, the real part is only fixed up to a real constant.
We make use of this freedom by subtracting the value at s=0
\ba
Re{\tilde G}_{\pi \pi}(s,k) & = & ReG_{\pi \pi}(s,k)-ReG_{\pi \pi}(0,k)
\nonumber \\
 & = & -\frac{1}{\pi}{\cal P} \int ds' \ \frac{ ImG_{\pi \pi}(s,k)}
{s-s'} \ \frac{s}{s'}.
\ea
The resulting, once subtracted, 2-pion propagator takes the form
\beq
{\tilde G}_{\pi\pi}(s,k)=\frac{1}{\omega_k} \ \frac{1}
{s-4\omega_k^2+i\eta} \ \frac{s}{4\omega_k^2} \ .
\eeq
which differs from the non-subtracted one (compare
eq.~(25)) by a factor $s/4\omega_k^2$. Notice that the imaginary part
remaines unchanged, i.e.
\beq
Im G_{\pi \pi}(s,k)=Im {\tilde G}_{\pi\pi}(s,k) \ ,
\eeq
thus maintaining the unitarity. Recalling the BbS T-matrix equation
(24) with the once-subtracted propagator
\beq
T_{\pi\pi}^{JI}(E;q,q') = V_{\pi\pi}^{JI}(E;q,q')
+ \int_0^{\infty}dk \ k^2 \ 4{\omega}_k^2 \
V_{\pi\pi}^{JI}(E;q,k)  \   G_{\pi\pi}^0(E,k) \
\frac{s}{4\omega_k^2} \  T_{\pi\pi}^{JI}(E;k,q')  ,
\eeq
one realizes that the subtraction factor $s/4\omega_k^2$ cures the
divergence of the scattering length since, in the soft pion limit,
$s \propto m_\pi^2$. By adjusting the form factors ($\Lambda_1= 1.5 GeV
,\Lambda_2= 3 GeV ,\Lambda_3= 4 GeV$)
we obtain $a^{00}=0.20m_\pi^{-1}$ and $a^{02}=-0.031m_\pi^{-1}$ which
are in perfect agreement with the most recent experimental
determinations \cite{BuLo}. As a result of the correct
chiral behaviour of the scattering length (which we also checked
numerically) the 'subtracted' M-amplitude again shows repulsion close
to E=0 (full line in the lower part of Fig.~10).
Above threshold, our fit underestimates the phase shifts (note
the different scale as compared to Fig.~9).
This latter point will be improved in the next section.

To end this section, let us turn to the in-medium
$\pi \pi$ M-amplitude calculated by means of the once subtracted 2-pion
propagator eq.~(51). In section 3 it was shown that $\pi\pi$
interactions without chiral constraints may lead to condensation.
In the linear sigma model
this was not the case as a consequence of a repulsion in the
M-amplitude below the $2m_\pi$ threshold. The same mechanism is at work
when applying a once subtracted dispersion relation to the
2-pion propagator of the scattering equation in the BbS framework.
As one can see from the upper part of Fig.~10, the repulsion in the
iterated M-amplitude prevents the instability.

%%%%%%%%%%%%%%%%%%%%%%%%%%%%%%%%%%%%%%%%%%%%%%%%%%%%%%%%%%%%%%%%
\subsection{The J\"{u}lich model with chiral constraints}
%%%%%%%%%%%%%%%%%%%%%%%%%%%%%%%%%%%%%%%%%%%%%%%%%%%%%%%%%%%%%%%%

To be more realistic in the description of the phase shifts
over a wide range of energies, we here present a preliminary version of
the J\"ulich model \cite{Lohs} including constraints
from chiral symmetry. As described in the previous section, we
impose the correct chiral limit for the scattering lengths
by applying the subtraction method (51) to both the 2$\pi$ and
the $K\bar K$ propagator. It turns out that the 1/${4\omega_k^2}$
dependence of the subtraction factor in (54) leads to
an appreciable suppression of the pseudopotentials $V^{JI}$ in the
integrand at small momenta, especially for the contribution from
t-channel $\rho$ exchange. This causes a considerable reduction of
the attraction in the (00) channel above threshold. We have therefore
added the 4-pion contact interactions generated by the
Weinberg Lagrangian as well as an $f_2(1270)$ t-channel exchange
in the $\pi\pi$ and in the $K\bar K$ system. The importance of
the latter has already been pointed out in ref.\cite{Bugg}.
The resulting fit to the phase shift is not quite as good as in
the original J\"ulich model, but it qualitatively reproduces the
important features in all partial waves up to J=2 and $E\approx 1.1 GeV$
(see Fig.~11 for the case of the JI=00 channel, which is the one of
interest in the present discussion).
The resulting in-medium $M^{00}$ amplitude reconfirms the findings of
the two previous sections. Even at very high densities
there is only a very moderate accumulation of strength slightly
below $2m_\pi$, i.e. the system
is far from condensation (see Fig.~12). The reason is the same as
before namely the real part of the $\pi\pi$ interaction
changes sign and becomes repulsive at energies below the
$2 m_\pi$ threshold. When the chiral constraints are imposed this
feature is preserved at the $T$-matrix level. Nevertheless, we observe
a resonance structure below threshold (insert in the upper panel of
Fig.~12), which corresponds to a quasi bound 2-pion state.

%%%%%%%%%%%%%%%%%%%%%%%%%%%%%%%%%%%%%%%%%%%%%%%%%%%%%%%%%%%%%%%%%%%%
\section{Summary and conclusion}
%%%%%%%%%%%%%%%%%%%%%%%%%%%%%%%%%%%%%%%%%%%%%%%%%%%%%%%%%%%%%%%%%%%%
Based on a conventional description of single-pion
propagation in the nuclear medium \cite{ErWe} and various models of the
vacuum $\pi\pi$ interaction we have studied s-wave $\pi\pi$ correlations
in symmetric nuclear matter. When using phenomenological interactions
which give a realistic description of the phase shifts in low partial
waves, we find $\pi^+\pi^-$ s-wave pair condensation with a critical
density only sightly above saturation density, thus confirming earlier
findings \cite{Aou93}. The reason is a strong
coupling to the ph continuum
combined with the attractive s-wave $\pi\pi$ interaction. Such an
instability would have dramatic consequences for the nuclear equation
of state.

One problem with phenomenological two-pion interactions is
that they do not obey the constraints imposed by chiral symmetry,
most noteably the correct behavior of the scattering lengths in
the chiral limit $m_\pi\to 0$. When constructing scattering potentials
from chirally symmetric Lagrangians (the linear $\sigma$ model and the
Weinberg model) which extrapolate correctly to zero pion mass, we have
shown that the instability disappears and only minor modifications
are obtained in the subthreshold region. These findings are generic
and do not depend on the details of the interaction. To be more
realistic,
we have also studied an improved J\"ulich model in which the correct
chiral limit is imposed by a once-subtracted dispersion relation and
which gives a satisfactory description of the phase shifts up to 1 GeV.
Also in this case instabilities are avoided.

In conclusion we want to argue, that the in-medium s-wave $\pi\pi$
correlations crucially depend on chiral symmetry constraints, especially
the correct chiral limit for the scattering length. Once these are
implemented in the theory, the density effects are rather small.

\vspace{2cm}
\begin{center}
{\bf Acknowledgement}
\end{center}
\vspace{0.5cm}

\noindent
We thank G. E. Brown, J. Durso and W. N\"orenberg for useful
discussions.
This work was supported in part by a grant from the National Science
Foundation, NSF-PHY-89-21025. One of us (R.R.) acknowledges financial
support from Deutscher Akademischer Austauschdienst (DAAD) under program
HSP II/AUFE.
\vfill\eject

\vfill\eject

\newpage
\begin{center}
{\Large \sl \bf Figure Captions}
\end{center}
\vspace*{1.5 cm}

\begin{itemize}
\item[{\bf Figure 1}:] Particle-hole and $\Delta$-hole bubbles.

\item[{\bf Figure 2}:] Dispersion equations for the most important
pionic
branches in nuclear matter at $\rho=\rho_0$. The solid curves designate
the particle-hole continuum. Dashed curves are for the 'interacting'
$\Delta$-hole and pion branches. Here the $\Delta$-hole and the pionic
branch are decoupled from the p-h continuum. For the in-medium nucleon
mass we take 80\% of the free one.

\item[{\bf Figure 3}:] $\sigma$ - strength function for $\rho=\rho_0$.
Dashed curve is for the pure two level model of \cite{Cha91}. The bound
state is situated, for this density, at $E_{\pi\pi}=1.92m_{\pi}$.
The solid curve is for a two level model with the corrections
mentioned in the text, namely the $\Delta$-decay width and
the 2p-2h pion self energy.

\item[{\bf Figure 4}:]
{\bf (a)} Imaginary part of the
$\sigma$-self energy in the pure two level model (dashed curves
for 0.5$\rho_0$ and $\rho_0$) and in the complete model (solid
curves for 0.5$\rho_0$, $\rho_0$ and 1.3$\rho_0$.

{\bf (b)} Real part of the inverse $\sigma$-propagator,
$D_{\sigma}^{-1}$, in the full model. The solid curve
is for $0.5\rho_0$, the long-dashed curve for $\rho_0$ and the
short-dashed curve for $1.3\rho_0$. As explained in the text,
the quasi pole appears for $\rho=\rho_0$ at $E_{\pi\pi}=1.9m_{\pi}$
and then is pushed to $E=0.2m_{\pi}$ at $\rho=1.3\rho_0$.

\item[{\bf Figure 5}:] Evolution of the $\sigma$-strength function
in the full model calculated with the separable $\pi\pi$ interaction.
At $0.5\rho_0$ and $\rho_0$, the imaginary
part begins and then invades the subthreshold region. For $1.3\rho_0$,
it is constituted in a resonant like structure around the quasi pole.
The latter condenses for $\rho>1.3\rho_0$.

\item[{\bf Figure 6}:] Lippmann-Schwinger equation for the $\pi\pi$
T-matrix in the meson exchange model of the J\"ulich group.
Below $E\approx 1GeV$ the dominant contribution is $\rho$ $t$-channel
exchange.

\item[{\bf Figure 7}:] $\sigma$-strength function for the
J\"ulich $\pi\pi$ interaction using the full model for the one-pion
propagator. Here the $\pi\pi$ bound state condenses at slightly
lower density than in the separable model, namely at $\rho\approx
1.3\rho_0$. The zero near 7$m_\pi$ is caused by the opening of the
$K\bar K$ channel.

\item[{\bf Figure 8}:] Fit to the $\pi\pi$ phase shifts in the JI=00
channel using the linear $\sigma$ model.

\item[{\bf Figure 9}:] In-Medium results for the imaginary part of the
invariant $M^{00}$ amplitude calculated in the linear $\sigma$ model.
Solid curve: vacuum result, long-dashed: $\rho=0.5\rho_0$,
short-dashed: $\rho=\rho_0$, dotted: $\rho=2\rho_0$.

\item[{\bf Figure 10}:] Invariant $M^{00}$ amplitude in the gauged
nonlinear $\sigma$ model (Weinberg Lagrangian) employing the subtraction
scheme: full line: vacuum results, long-dashed line: $\rho =0.5\rho_0$,
short-dashed line: $\rho=\rho_0$, dotted line: $\rho=2\rho_0$.

\item[{\bf Figure 11}:] Phase shift fit within the extended J\"ulich
model

\item[{\bf Figure 12}:] In-medium $M^{00}$ amplitude calculated with
the extended J\"ulich model:  \\
full line: vacuum results, long-dashed line: $\rho =0.5\rho_0$,
short-dashed line: $\rho=\rho_0$, dotted line: $\rho=2\rho_0$.
\end{itemize}

\end{document}